%% file: Ising_single_chain.tex
\documentclass[aps,prl,10pt,twocolumn,letterpaper,amsmath,amssymb]{revtex4-2}

\usepackage[utf8]{inputenc}
\usepackage{graphicx}
\usepackage{subfigure}
\usepackage[usenames,dvipsnames]{xcolor}
\usepackage{epstopdf}
\usepackage{amsmath,amsthm,amssymb}
\usepackage{latexsym}
\usepackage{bm}
\usepackage{dcolumn}
\usepackage{braket}
\usepackage[normalem]{ulem}
\usepackage{times}
\usepackage{hyperref}
\usepackage{enumitem}
\newcommand{\ignore}[1]{}

\newcommand{\red}[1]{\textcolor{black}{#1}}

\newcommand{\blue}[1]{\textcolor{black}{#1}}

\newcommand{\mytitle}{Phase Switch Driven by the Hidden Half-Ice, Half-Fire State in a Ferrimagnet}

\begin{document}
%\begin{CJK*}{}{} % Use default fonts from CJK
%\bibliographystyle{apsrev4-1}
%\bibliographystyle{science}
%\bibliographystyle{naturemag}

\title{\mytitle}
\author{Weiguo Yin}
\email{wyin@bnl.gov}
\author{A. M. Tsvelik}
\affiliation{Condensed Matter Physics and Materials Science Division,
Brookhaven National Laboratory, Upton, New York 11973, USA}

\date{\today}

\begin{abstract}
The notion of ``half fire, half ice'' was recently introduced to describe an exotic macroscopic ground-state degeneracy emerging in a ferrimagnet under the critical magnetic field, in which the ``hot'' spins are fully disordered on the sublattice with smaller magnetic moments and the ``cold'' spins are fully ordered on the sublattice with larger magnetic moments. %which separates the ferromagnetic and antiferromagnetic phases in the ground-state phase diagram.
Here we further point out that this state has a twin named ``half ice, half fire'' in which the hot and cold spins switch positions. The new state is an excited state---thus hidden in the ground-state phase diagram---and \red{is robust with respect to the} %it can survive 
interactions that destroy the half fire, half ice state. We demonstrate with exact results how this hidden state can drive phase switching at desirable finite temperature, even for the one-dimensional Ising model where phase transition at finite temperature is forbidden. We suggest that our findings may open a new door to the understanding and controlling of phase competition and transition in unconventional frustrated systems.

%\bigskip
\end{abstract}

%PhySH:
%Interdisciplinary Physics => Chemical Physics \& Physical Chemistry => Thermodynamics
%Statistical Physics => Classical statistical mechanics
%Discipline(s) => Physical System(s) => Concept(s)
%Condensed Matter \& Materials Physics, Statistical Physics => Many-body techniques

\maketitle
%\end{CJK*}

%\section{Introduction}
%More than 12,000 papers have been published between 1969 and 1997 using the Ising model.

\emph{Introduction}---\red{Finding new states with exotic physical properties and understanding/controlling their phase transitions are central problems in condensed matter physics and materials science. Such states are abound} in frustrated magnets %which give rise to such exotic magnetic states 
as spin ice, spin glass, spin liquid, \red{and skyrmion} that may play important roles in quantum computing, spintronics, and unconventional superconductivity~\cite{Balents_nature_frustration,Miyashita_10_review_frustration,NNano_13_review_skyrmion}\red{, and are a fundamental test ground for applications of machine learning to complex systems~\cite{Fan_NC_23_ML-2D-Ising-glass_ML,Roth_PRB_23_2D-J1-J2_ML-VMC}}. The essence of frustration is usually assumed to be the ground-state degeneracy that emerges as a consequence of competing exchange interactions among the spins. %Such geometrical frustration demands 
%It is usually assumed that in frustrated systems some or all of the exchange interactions must be either antiferromagnetic (AF) or %%that in case the exchange interactions are all ferromagnetic (FM), they 
%strongly direction-dependent such as in the Kitaev model~\cite{Kitaev_2006}. 
\red{Therefore, frustration is also widely invoked to explain rich phase diagrams %and phase transitions, a hallmark 
of strongly correlated systems~\cite{Dagotto_Science_review}.} %such as high-$T_c$ superconductors, manganites with colossal magnetoresistance, twisted bilayer graphene, etc. 

\red{In this paper, we explore the frustration-driven mechanism of phase switch at finite temperature.} \blue{Its design %of a frustration-driven (or entropy-driven) 
has usually been carried out by first computing the zero-temperature phase diagram, in which two phases with frustration-induced macroscopically different degeneracy have their own stable regions, and then placing the system near their phase boundary and on the side with less entropy.} \red{When heated up, the system will enter the other phase with lower free energy gained from high entropy~\cite{Miyashita_10_review_frustration,016_Strecka_book_chapter}. Here, we study an alternative mechanism, where the sharp phase switch is driven by the high entropy of an excited state that is \emph{hidden} in the zero-temperature phase diagram.}

\begin{figure}[tb]
    \begin{center}
\includegraphics[width=0.99\columnwidth,clip=true,angle=0]{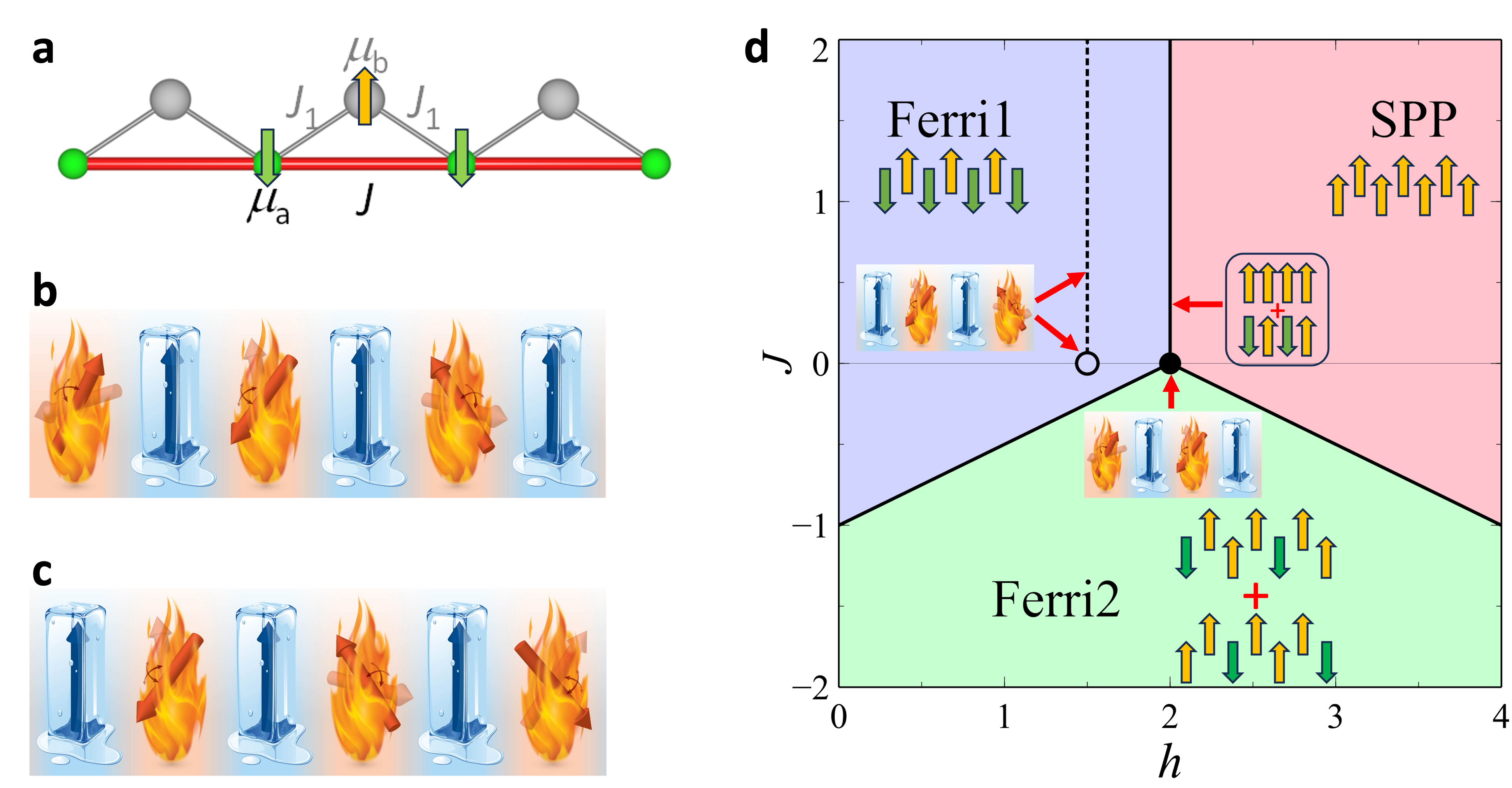}
    \end{center}
\caption{Model, exotic states, and the zero-temperature phase diagram. (a) A schematic diagram of the minimal model, which consists of the original Ising spins (green balls) coupled by the FM interaction $J>0$ (red bonds) and the decorated Ising spins (gray balls) coupled to the backbone by the AF interaction $J_1<0$ (gray bonds). Their magnetic moments are $\mu_a$ and $\mu_b$, respectively. The arrows pointing up or down means the spin value of $+1$ or $-1$. (b) A cartoon illustration of the half-fire, half-ice state with fully disordered and fully ordered spins. (c) A cartoon illustration of the half-ice, half-fire state in which the hot and cold spins switch positions. (d) The zero-temperature phase diagram in the $h-J$ plane of the minimal model for $J_1=-1$, $\mu_a=1$, and $\mu_b=4/3$. The shorthand notation: Ferri1 - the standard ferrimagnetic phase, Ferri2 - the second ferrimagnetic phase, SPP - the saturated paramagnetic phase. The solid circle is the critical point hosting the half-fire half ice state, which does not survive finite $J$ \red{because its macroscopic degeneracy is lifted}. The open circle and dashed line host the hidden half-ice half fire excited state, \red{whose  macroscopic degeneracy is not lifted} as $J$ increases, forming the boundary line between two excited states inside the Ferri1 phase: SPP and inverse Ferri1 where all the spins in Ferri1 flip.}
\label{Fig:structure}
\end{figure}

Recently, there has been an intense search for frustrated magnetism beyond the ``standard model'' of condensed matter physics~\cite{vandenBrink2011,Smerald2015,Kitaev_2006,Yao-Lee_PRL_2011_Kitaev_SU2,PRL_17_spiral_DE,Yin_PRB_21_trimer,Yin_PRL_22_trimer}. One of the intriguing results is the macroscopic ground-state degeneracy of $2^N$ (where $2N$ is the total number of the spins) emerging in an Ising ferrimagnet under the critical magnetic field, in which the spins on the sublattices with smaller and larger magnetic moments are fully disordered and fully ordered, respectively---thus noted as ``half fire, half ice'' (HFHI) [Fig.~\ref{Fig:structure}(b)]---even in the absence of the aforementioned conventional geometrical frustration~\cite{Yin_g}. This exotic state can help transform the $XXZ$ quantum spin model to the $XY$ model. The partial fire, partial ice phenomenon was anticipated to generally exist in systems that could be modeled as ferrimagnets in magnetic field~\cite{Yin_g} and subsequently exhibited in Ising-Heisenberg ferrimagnets~\cite{Torrico_g1,Torrico_g2}. It is thus timely to find out how the fire-ice phenomenon can be used to generate and control competition and switches among different phases at not only zero temperature but also more accessible finite temperature. \red{Yet, it is challenging to implement this with the HFHI state in the aforementioned traditional way, as the HFHI state is a fine-tuned point in the zero-temperature phase diagram [solid circle in Fig.~\ref{Fig:structure}(d)] and disappears with small changes in the model parameters.}  

The purpose of this Letter is to reveal that the HFHI state has a ``half ice, half fire'' (HIHF) twin, in which the hot and cold spins switch positions [Fig.~\ref{Fig:structure}(c)]. The new state has the same macroscopic degeneracy of $2^N$; however, it is always an excited state \red{(thus hidden) in the zero-temperature phase diagram. It does have the merit of being robust with respect to the interaction [open circle and the dashed line in Fig.~\ref{Fig:structure}(d)] that otherwise energetically disfavors the HFHI state. We then ask whether} it is possible to utilize this hidden frustration to drive phase changes at finite temperature. Below  we attempt to get an unambiguous insight into this effect using an exact model example. We note that this task is nontrivial since to date even the simple two-dimensional (2D) Ising model in a magnetic field has not been solved ~\cite{Mattis_book_08_SMMS}, while finite-temperature phase transition is forbidden in 1D Ising models with short-range interactions~\cite{Ising1925}. \red{In addition, the previous models of utilizing frustration to yield sharp phase crossover in the 1D Ising models suffer from the feature that the crossover temperature goes to zero as the crossover gets narrower~\cite{016_Strecka_book_chapter,017_Krokhmalskii_PA_21_3-previous-chains_effective_model}.}
Very recently, it was discovered that the forbidden transition in the 1D Ising model can be approached arbitrarily closely at fixed finite temperature $T_0$ in decorated ladder~\cite{Yin_MPT} and single chain~\cite{Yin_MPT_chain} by \red{independently} making the crossover width $2\delta T$ exponentially narrower and narrower ($\delta T=0$ means a genuine transition), resulting in an ultranarrow phase crossover (UNPC) with large latent heat that would be characterized as a genuine first-order phase transition in a routine laboratory measurement. \red{Such UNPC was established solely based on analysis of the mathematical structure of the first-order phase transition by mimicking a nonanalytic function with an analytic function, implying a wide applicability of the UNPC notion and calling for further insights into the utility of UNPC in studying various outstanding physical problems.} This inspires us to explore the HIHF-driven phase switch via UNPC in a minimal model that can be easily generalized. 

\emph{The model}---We consider a minimally decorated single-chain Ising model given by $H=H_a+ H_b$, where
\begin{eqnarray}
H_a&=&-J\sum_{i=1}^{N}\sigma_{i}\sigma_{i+1}-h\mu_a \sum_{i=1}^{N}\sigma_{i}, \label{ordinary} \\
H_b&=&-J_1\sum_{i=1}^{N}(\sigma_{i}+\sigma_{i+1}) b_i- h\mu_b \sum_{i=1}^{N}b_i.
\label{minimal}
\end{eqnarray}
As shown in Fig.~\ref{Fig:structure}(a), $H_a$ describes the backbone of the single chain with $\sigma_{i}=\pm1$ (green balls, referred to as \emph{type-a} spins) and $J>0$ the FM interaction (red bonds). This is the model originally studied by Ernst Ising~\cite{Ising1925}. $H_b$ describes the decorated parts, where $b_i=\pm1$ (gray balls, referred to as \emph{type-b} spins) bridges the $i$th and $(i+1)$th type-\emph{a} spins with the AF interaction $J_1<0$. $h$ depicts the magnetic field, $\mu_a$ and $\mu_b$ the magnetic moments of type-\emph{a} and type-\emph{b} spins, respectively. The relationship of $\mu_b > \mu_a > 0$ is used to represent ferrimagnetism~\cite{note:ferrimagnet}. $N$ is the total number of the unit cell and $\sigma_{N+1}\equiv\sigma_{1}$, $b_{N+1}\equiv b_{1}$ (i.e., the periodic boundary condition). It is crucial to note that this minimal model with only one decorated Ising spin per unit cell contains no geometric frustration in the sense that the AF $J_1$ and the FM $J$ interactions can be satisfied simultaneously by a proper arrangement of the spins.

\emph{Zero-temperature phase diagram}---As shown in Fig.~\ref{Fig:structure}(d), the zero-temperature phase diagram in the $h-J$ plane of the present minimal model contains three stable phases but none of them has macroscopic degeneracy; therefore, the crossover between them as temperature changes will be broad if any. The two special cases of (1) $\mu_a=\mu_b$, both $J<0$ and $J>0$~\cite{Stephenson_CanJP_70_J1-J2-Ising-chain}, and (2) $\mu_a < \mu_b$, $J=0$ ~\cite{Yin_g,Bell_JPC_74_Ising_ferri_1} were examined before and no sign of UNPC was found even for AF $J<0$ with geometric frustration. Nevertheless, the HFHI critical point (black solid circle) was found for $J=0$ and $h=h_c$ where $h_c \equiv 2(-J_1)/\mu_a$~\cite{Yin_g}. 
%By contrast, the HIHF state at $h=h_f < h_c$ where $h_f \equiv 2(-J_1)/\mu_b$ (open circle and dashed line) is the boundary between two excited states, thus hidden in the zero-temperature phase diagram. 

\begin{figure*}[th]
    \begin{center}
%        \subfigure[][]{
\includegraphics[width=0.95\textwidth,clip=true,angle=0]{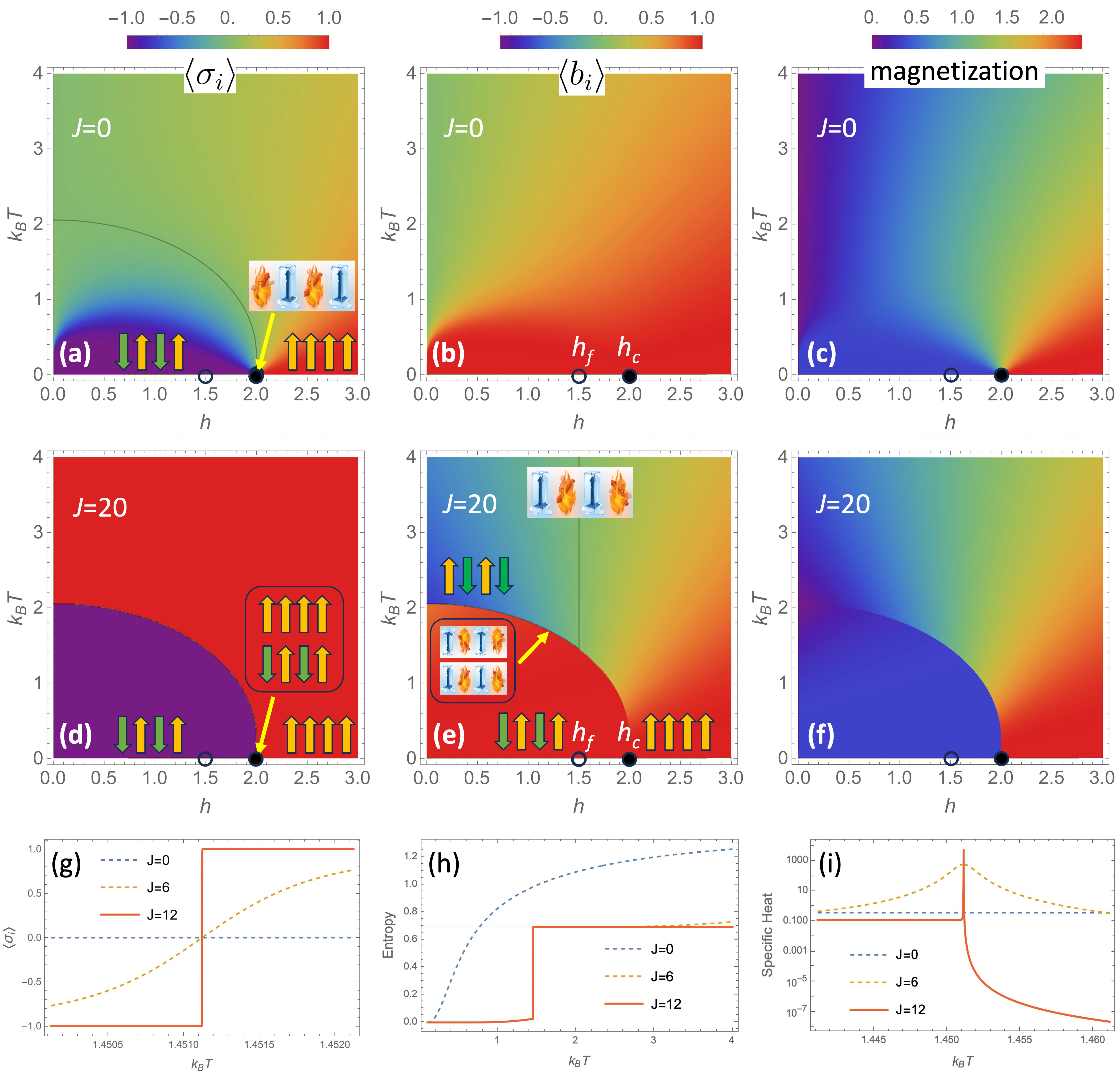}
%        }
    \end{center}
\caption{The density plots of $\langle\sigma_i\rangle$, $\langle \red{b}_i\rangle$, and the total magnetization $\mu_a\langle\sigma_i\rangle+\mu_b\langle b_i\rangle$ in the $h - T$ plane. Top panels: (a)-(c) for $J=0$. Bottom panels: (d)-(f) for $J=20$. The other model parameters are $J_1=-1$, $\mu_a=1$, and $\mu_b=4/3$, resulting in $h_c=2$ \red{(solid circle)} and $h_f=1.5$ \red{(open circle)}. The black line in (a),(d),(e) denotes the contour of zero value: In (a),  the $\langle\sigma_i\rangle=0$ line defines $T_0$ the phase boundary in (d) by Eq.~(\ref{eq:T0}); the yellow arrow points to the zero-temperature critical point at $h=h_c$ hosting the half-fire, half-ice state. In (e), the vertical line $\langle b_i\rangle=0$ at $h=h_f$ and above the phase boundary line hosts the opposite half-ice, half fire state, originates from the hidden frustration, and divides the region above $T$=$T_0$ in half where the decorated spins flip for $h<h_f$. The temperature dependence of (g) $\langle\sigma_i\rangle$, (h) entropy, and (i) specific heat for several $J$ at $h=h_f$.}
\label{Fig:h-T}
\end{figure*}

\emph{Ultranarrow phase crossover at finite temperature}---We describe the mathematical details in the Supplemental Material~\cite{SI} and show key results below. The order parameter that allows for fast and accurate identification of UNPC is the magnetization of the type-\emph{a} spins for $h\ne 0$ [see Supplemental Equation (S7)] given by, 
\begin{equation}
    \langle \sigma_i \rangle=-\frac{\partial f}{h\,\partial \mu_a}, 
\end{equation} 
which switches sign at $T_0$ with the crossover width $2\delta T_0 = 2\left(\frac{\partial \langle \sigma_i \rangle}{\partial T}\right)^{-1}_{T=T_0}$ [see Supplemental Equation (S8)]. Now take a close look at Fig.~\ref{Fig:h-T}(a) for $J=0$, there is the $\langle \sigma_i \rangle=0$ line and $\langle \sigma_i \rangle$ changes sign as $T$ crosses the line, which determines $T_0$ by
\begin{equation}
e^{4\beta_0 J_1} =\frac{\sinh[\beta_0 h(\mu_b - \mu_a)]}{\sinh[\beta_0 h(\mu_b + \mu_a)]}, \label{eq:T0}
\end{equation}
where $\beta_0=1/(k_\mathrm{B}T_0)$ with $k_\mathrm{B}$ being the Boltzmann constant. Eq.~(\ref{eq:T0}) has a real solution for $\mu_b > \mu_a > 0$, $J_1<0$, and $|h|< h_c$. Clearly, $T_0$ is independent of $J$.
The problem is that the crossover width $2\delta T(J)$ is too broad for $J=0$ [Fig.~\ref{Fig:h-T}(a)], but it can be exponentially narrowed by $J>0$ in the form of $2\delta T(J)=e^{-2\beta_0 J} 2\delta T(J=0)$, as shown in Fig.~\ref{Fig:h-T}(d), resulting in an UNPC at $T_0$ from $\langle \sigma_i \rangle=-1$ to $+1$.

What is the underlying mechanism of this unexpected UNPC? As shown in Fig.~\ref{Fig:h-T}(h),
a sharp jump in entropy of about $k_\mathrm{B}\ln 2$ per unit cell occurs within the ultranarrow crossover for $h=h_f$ where $h_f\equiv 2(-J_1)/\mu_b=h_c\mu_a/\mu_b$, resembling a genuine first-order phase transition. This means that one spin per unit cell is decoupled from the spin system and the field. In the zero-temperature HFHI state at $h=h_c$ [the solid circle in Fig.~\ref{Fig:structure}(d) and pointed by the yellow arrow in Fig.\ref{Fig:h-T}(a)], the type-\emph{b} spins are fully aligned by the field due to $|\mu_b|>|\mu_a|$ but the type-\emph{a} spins are decoupled from the spin system and the field~\cite{Yin_g}. However, this exotic state does not survive finite $J$. Now we consider the HIHF state at $h=h_f<h_c$ where the hot and cold spins switch positions, i.e., type-\emph{a} spins are fully aligned by the field together with the strong $J>0$ but the type-\emph{b} spins are decoupled from the spin system and the field [the open circle at $h_f$ in Figs.~\ref{Fig:structure}(d) and \ref{Fig:h-T}(e)]. This new exotic state is an excited state everywhere in the zero-temperature phase diagram and is thus hidden. Yet, \red{its  macroscopic
degeneracy is not lifted by}  $J>0$, forming the boundary line between two excited states [the dashed line in Fig.~\ref{Fig:structure}(d)], and emerges as the driving force of the new type of frustration-driven UNPC. 

\begin{figure*}[th]
    \begin{center}
%        \subfigure[][]{
\includegraphics[width=0.95\textwidth,clip=true,angle=0]{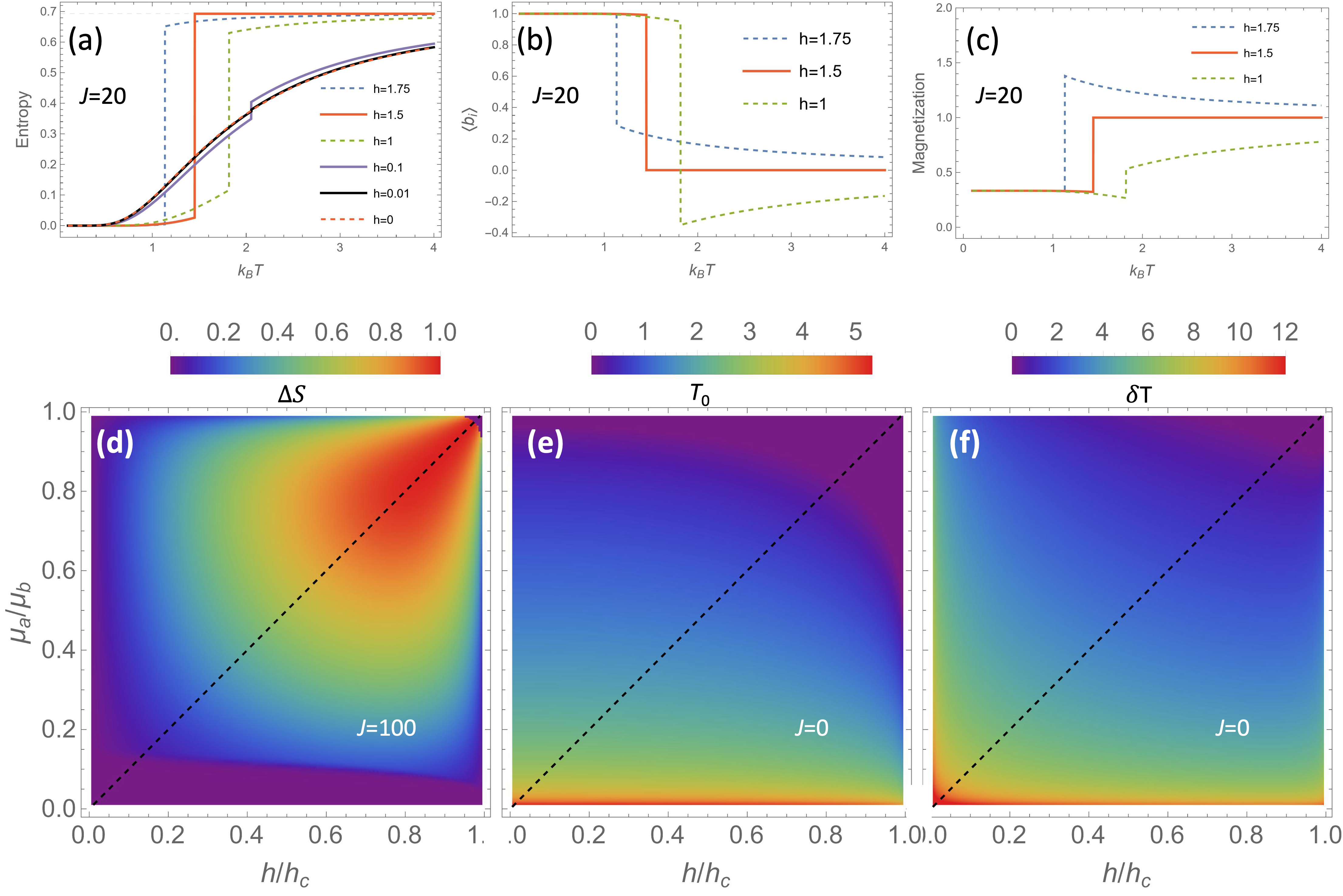}
%        }
    \end{center}
\caption{The temperature dependence of (a) entropy, (b) $\langle b_i\rangle$, and (c) the total magnetization $\mu_a\langle\sigma_i\rangle+\mu_b\langle b_i\rangle$ for $\mu_b=4/3$, $J=20$ and several $h$'s. The density plots in the $h/h_c - \mu_a/\mu_b$ plane of (d) $\Delta S$ the entropy jump at $T_0$ in the unit of $k_\mathrm{B}\ln 2$ per unit cell for $J=100$, (e) $T_0$, and (f) $\delta T$ for $J=0$. The other model parameters are $J_1=-1$ and $\mu_a=1$; $h_c\equiv 2(-J_1)/\mu_a=2$. }
\label{Fig:h-mu}
\end{figure*}

This mechanism is clearly demonstrated by the exotic behavior of $\langle b_i \rangle= -\frac{\partial f}{h\,\partial \mu_b}$ the magnetization of the type-\emph{b} spins. For $J=0$, $\langle b_i \rangle$ is always positive, following the magnetic field due to $\mu_b > \mu_a$ [Fig.~\ref{Fig:h-T}(b)]. For sufficiently large $J>0$, outside the ultranarrow crossover region, 
\begin{equation}
    \langle b_i \rangle  \approx\left\{
\begin{array}{lcc}
  \tanh[\beta(h \mu_b - 2 J_1)] > 0\; &\mathrm{for}& \; T < T_0. \\
  \tanh[\beta(h \mu_b + 2 J_1)]\; &\mathrm{for}& \;\;\; T > T_0, \\
\end{array}
\right. \label{OP_s}
\end{equation}
where $\beta=1/(k_\mathrm{B}T)$. For $T>T_0$, there are three distinct regions for the UNPC [Figs.~\ref{Fig:h-T}(e) and \ref{Fig:h-mu}(b)] corresponding to the three excited states inside the regime where Ferri1 is the ground state shown in Fig.~\ref{Fig:structure}(d): \blue{(i) $0<h<h_f$} where $ \langle b_i \rangle <0$, i.e., both the type-\emph{a} and type-\emph{b} spin flip from the Ferri1 ground state at the UNPC. Note that the flipping of the type-\emph{b} spins  means the existence of the $ \langle b_i \rangle =0$ line within the ultranarrow crossover. 
\blue{(ii) $h=h_f$} where $ \langle b_i \rangle =0$ and the type-\emph{a} spins flip but the type-\emph{b} spins are decoupled from the spin system and the field (inferred together with information about the entropy jump $\Delta S$ of about $k_\mathrm{B}\ln 2$ per unit cell), i.e., the HIHF state. \blue{(iii) $h_f<h<h_c$} where $ \langle b_i \rangle >0$, i.e., the type-\emph{a} spins flip but the type-\emph{b} spins do not at the UNPC, which is \red{seemingly} SPP like. \red{However, it has much stronger $\langle \sigma_i \rangle$ and much weaker $\langle b_i \rangle$ than the SPP-like phase above $T_0$ for $J=0$. More quantitative data on} the temperature evolution of entropy, $ \langle b_i \rangle$, and the system's total magnetization per unit cell $ m= -\frac{\partial f}{\partial h}=\mu_a\langle \sigma_i \rangle+\mu_b\langle b_i \rangle$ in these three regions are shown in Figs.~\ref{Fig:h-mu}(a)-\ref{Fig:h-mu}(c). \red{It is clear that the entropy jumps and $\langle b_i \rangle$ drops a lot even for $h>h_f$, indicating that most decorated spins are ``on fire'' (i.e., decoupled from the other spins). Therefore, the (partial) HIHF state is still the main driving force of the UNPC for $h>h_f$. It would be interesting to illustrate in subsequent publications the evolution of the HIHF state such as the redistribution of the spins on fire, especially for going on to higher dimensions where the exact solutions are not yet available.}

The density plots of $\Delta S$, $T_0$, and $\delta T(J=0)$ in the $h/h_c-\mu_a/\mu_b$ plane are shown in Figs.~\ref{Fig:h-mu}(d)-\ref{Fig:h-mu}(f), respectively. The largest value of $\Delta S$ is along the $h/h_c=\mu_a/\mu_b$ line, which manifests that the hidden HIHF frustrated state developed from $h=h_f$ and $T=0$ is indeed the driving force of the above UNPC, which resembles the genuine first-order phase transition with large latent heat. 
Besides, it shows that this type of UNPC is difficult to be realized for small $\mu_a/\mu_b$ or weak $h$. We can avoid small $\mu_a/\mu_b$ but weak $h$ is relevant to device application, so it deserves further investigation. 

In summary, we have used rigorous methods to establish  the existence of an exotic half ice, half fire state and its effects on driving phase switch in a minimal 1D model for ferrimagnets in magnetic field. The phase switch is identified as ultranarrow phase crossover, which is anticipated to become a genuine phase transition in higher dimensions. Thanks to its simple structure, the minimal model may be attractive for device applications. Since various physical Ising models have already been implemented in electronic circuits~\cite{Ising_FPGA}, optical %neural 
networks~\cite{Pierangeli_IsingMachine_PRL19}, and optical lattice~\cite{Bernien_Nature_17_Rydberg}, %\cite{Inagak_IsingMachine_Science16,Inagak_IsingMachine_Science16,Pierangeli_IsingMachine_PRL19,Yamamoto_npjQI_Ising}.  
the making of the HIHF-driven UNPC-based 1D devices right away seems to be feasible; \red{potential applications could range from temperature sensing to refrigerating based on the magnetocaloric effect~\cite{PRXEnergy_24_magnetocaloric}. That $T_0$ and $2\delta T$ are independently controlled by different model parameters is a highly desirable feature for application.} The model can be easily generalized \red{in both one and higher dimensions} (see Supplemental Fig. S1); therefore the ways to realize the effects of the ice-fire and fire-ice twin states are abundant. Further exploration of how those effects function in more complex systems---when spins become quantum and when they are coupled to charge, orbital, and lattice degrees of freedom---would be instrumental in the search for quantum materials for quantum information science and microelectronics. \red{Last but not least, the 1D systems with UNPC provide an easier test ground for applications of machine learning to frustrated magnets beyond zero temperature.}
\begin{acknowledgments}
%W.Y. is grateful to D. C. Mattis for mailing him a copy of Ref.~\cite{Mattis_book_08_SMMS} as a gift and the inspiration over the years. 
Our research at Brookhaven National Laboratory was supported by U.S. Department of Energy (DOE) Office of Basic Energy Sciences (BES) Division of Materials Sciences and Engineering under contract No. DE-SC0012704.
\end{acknowledgments}

%\bibliography{Ising_quantum}
\input{trilogy_3.bbl}

%\vspace{0.5cm}
%\noindent\textbf{Author Information} The author declares no competing interests. Correspondence and requests for materials should be addressed to W.Y. (wyin@bnl.gov).

%\vspace{0.5cm}
%\noindent\textbf{Data} Wolfram Mathematica version 13.2.0.0 was used to assist the derivation and computation. Numerical calculations with the 200-digit precision were performed to warrant the correct results.

%\newpage
%\beginsupplement
%\section*{[Supplementary Information] \mytitle}
%\input{SI}

\end{document}

%% file: trilogy_3.bbl
%apsrev4-2.bst 2019-01-14 (MD) hand-edited version of apsrev4-1.bst
%Control: key (0)
%Control: author (8) initials jnrlst
%Control: editor formatted (1) identically to author
%Control: production of article title (0) allowed
%Control: page (0) single
%Control: year (1) truncated
%Control: production of eprint (0) enabled
%

%% file: Ising_single_chain.bbl
\begin{thebibliography}{33}
\bibitem [{not()}]{note:ferrimagnet}%
  \BibitemOpen
  \href@noop {} {}\bibinfo {note} {The model is invariant under the transformations that satisfy $\mu_a \mu_b J_1 < 0$ and $|\mu_b| > |\mu_a|$. For example, $\mu_b=-3$ and $\mu_a=2$ with FM $J_1>0$ were recently realized in the mixed $3d-5d$ copper-iridium oxide Sr$_3$CuIrO$_6$~\cite{Yin_g,Yin_PRL_Sr3CuIrO6}).}\BibitemShut {Stop}%
\end{thebibliography}
